\documentclass[%
aip,
pop,
 amsmath,amssymb,
 reprint,%
]{revtex4-1}

\usepackage[utf8x]{inputenc}
\usepackage{color}
\usepackage{amsmath}
\usepackage{amssymb}
\usepackage{amsfonts}
\usepackage{amsthm}
\usepackage{float}
\usepackage{bm}
\usepackage{latexsym}
\usepackage{upgreek}
\usepackage{hyperref}
\usepackage{multirow}
\usepackage[capitalize]{cleveref}
\usepackage{siunitx}
\usepackage{graphicx}
\hypersetup{
    colorlinks,
    citecolor=blue,    filecolor=blue,
    linkcolor=blue,     urlcolor=blue
}

\begin{document}

%\preprint{APS/123-QED}

\title{Optimization and stabilization of a kilohertz laser-plasma accelerator}

\author{L.~Rovige}
\affiliation{Laboratoire d'Optique Appliqu\'ee, ENSTA, CNRS UMR7639, Ecole Polytechnique, Chemin de la Huni\`ere,91761 Palaiseau, France.}

\author{J. Huijts}%
 \affiliation{ LOA, CNRS, \'Ecole Polytechnique, ENSTA Paris, Institut Polytechnique de Paris, Palaiseau, France}
 
\author{I.A.~Andriyash}
\affiliation{Laboratoire d'Optique Appliqu\'ee, ENSTA, CNRS UMR7639, Ecole Polytechnique, Chemin de la Huni\`ere,91761 Palaiseau, France.}

 \author{A.~Vernier}
\affiliation{Laboratoire d'Optique Appliqu\'ee, ENSTA, CNRS UMR7639, Ecole Polytechnique, Chemin de la Huni\`ere,91761 Palaiseau, France.}

 \author{M.~Ouillé}
\affiliation{Laboratoire d'Optique Appliqu\'ee, ENSTA, CNRS UMR7639, Ecole Polytechnique, Chemin de la Huni\`ere,91761 Palaiseau, France.}

\author{Z.~Cheng}
\affiliation{Laboratoire d'Optique Appliqu\'ee, ENSTA, CNRS UMR7639, Ecole Polytechnique, Chemin de la Huni\`ere,91761 Palaiseau, France.}

\author{T.~Asai}
\affiliation{Graduate School of Maritime Sciences, Kobe University, Kobe, Japan}
\affiliation{ Kansai Photon Science Institute (KPSI), National Institutes for Quantum and Radiological Science and Technology (QST), Kyoto, Japan}

\author{Y.~Fukuda}
\affiliation{ Kansai Photon Science Institute (KPSI), National Institutes for Quantum and Radiological Science and Technology (QST), Kyoto, Japan}

\author{V. Tomkus}
\affiliation{Center for Physical Sciences and Technology, Savanoriu Ave. 231, LT-02300, Vilnius, Lithuania}
\author{V. Girdauskas}
\affiliation{Center for Physical Sciences and Technology, Savanoriu Ave. 231, LT-02300, Vilnius, Lithuania}
\affiliation{Vytautas Magnus University, K.Donelaicio St. 58. LT-44248, Kaunas, Lithuania}
 \author{G. Raciukaitis}
 \affiliation{Center for Physical Sciences and Technology, Savanoriu Ave. 231, LT-02300, Vilnius, Lithuania}
 \author{J. Dudutis}
 \affiliation{Center for Physical Sciences and Technology, Savanoriu Ave. 231, LT-02300, Vilnius, Lithuania}
 \author{V. Stankevic}
 \affiliation{Center for Physical Sciences and Technology, Savanoriu Ave. 231, LT-02300, Vilnius, Lithuania}
 \author{P. Gecys}
 \affiliation{Center for Physical Sciences and Technology, Savanoriu Ave. 231, LT-02300, Vilnius, Lithuania}

\author{R.~Lopez-Martens}
\affiliation{Laboratoire d'Optique Appliqu\'ee, ENSTA, CNRS UMR7639, Ecole Polytechnique, Chemin de la Huni\`ere,91761 Palaiseau, France.}

\author{J.~Faure}
\affiliation{Laboratoire d'Optique Appliqu\'ee, ENSTA, CNRS UMR7639, Ecole Polytechnique, Chemin de la Huni\`ere,91761 Palaiseau, France.}

\date{\today}
\begin{abstract}
Laser plasma acceleration at kilohertz repetition rate has recently been shown to work in two different regimes, with pulse lengths of either 30 fs or 3.5 fs. We now report on a systematic study in which a large range of pulse durations and plasma densities were investigated through continuous tuning of the laser spectral bandwidth. Indeed, two LPA processes can be distinguished, where beams of the highest quality, with 5.4 pC charge and a spectrum peaked at 2-2.5 MeV are obtained with short pulses propagating in moderate plasma densities. Through Particle-in-Cell simulations the two different acceleration processes are thoroughly explained.  Finally, we proceed to show the results of a 5-hour continuous and stable run of our LPA accelerator accumulating more than $18\times10^{6}$ consecutive shots, with 2.6\,pC charge and peaked 2.5\,MeV spectrum. A parametric study of the influence of the laser driver energy through PIC simulations underlines that this unprecedented stability was obtained thanks to micro-scale density gradient injection. Together, these results represent an important step towards stable laser-plasma accelerated electron beams at kilohertz repetition rate. 
\end{abstract}

\maketitle

\section{\label{sec:intro}Introduction}

Laser wakefield acceleration is a process in which an ultrashort laser pulse is focused into a plasma in order to drive a large amplitude plasma wave. Plasma electrons can be trapped and accelerated to relativistic energies in the giant electric fields ($\sim 100$~GV/m) of the plasma wave, resulting in compact acceleration. From the first proposals  \cite{taji79,gorb87,pukh02} and  demonstrations \cite{mode95,umst96,malk02,faur04,gedd04,mang04} to present days, laser-plasma accelerators (LPA)  have undergone tremendous progress following the constant improvements of laser technology. Two approaches have been recently pursued. The first aims at exploring the high power regime, where the laser power reaches the PW. Such lasers recently allowed the acceleration of electrons to multi-GeV energies \cite{HTKim:PRL2013,wang13,leem14} making them relevant drivers for compact sources of hard x-rays and $\gamma$-radiation, or future colliders \cite{Schroeder:PRSTAB2010}. Heat-management and optical damage in the high-power beamlines have been the main limitations to increasing the repetition rate beyond the 1-10 Hz range. The second approach focuses on the development of high-repetition rate sources that are of paramount importance for applications. A LPA running at kHz or faster will provide more stable beams with higher average currents and will permit the implementation of active feed-back loops to further stabilize the performance of the accelerator. Indeed, existing multi-mJ kHz laser systems are now able to provide laser pulses compressed to almost single-cycle wave packets, and with tight focusing, intensities in excess of $I\approx\SI[per-mode=symbol]{1e18}{\watt\per\square\cm}$ can be obtained, which is suitable for driving a LPA \cite{he13,beau15,salehi17,guenot17,faure19}. Due to such tight focusing, the acceleration length is typically short (few tens of micrometers), and the LPA electron energies are on the order of a few MeVs. Operation at kHz repetition rate provides a high average current (up to tens of nA), which makes them of a great interest for ultrafast electron diffraction \cite{he13b,he16,faure16}, irradiation experiments for radiation harness assessment \cite{hidd17} or radio-biology \cite{Rigaud2010,Lundh2012}. 

Recently, laser-plasma acceleration with kHz lasers was demonstrated in two different regimes: (i) one with 30~fs laser pulses of peak power of $P_0\approx 300$~GW and high plasma density $n_e>\SI{4e20}{\per\square\cubic\centi\meter}$\ \ \cite{salehi17}, and (ii) another with laser pulses that were post-compressed to nearly single-cycle durations $\tau=3.5$~fs, and reaching $P_0\gtrsim 420$~GW, but operating at lower plasma density \cite{guenot17,gustas18}. Both experiments resulted in the acceleration of MeV electrons but the accelerated beams had very different characteristics. Therefore, there is a need for an in-depth study of the effect of laser and plasma parameters in order to optimize the performance of kHz laser-plasma accelerators. 

In this paper, we report on a systematic study of the various acceleration regimes at play over a large range of plasma densities and pulse durations, allowing to understand the transitions between the different regimes. Our set-up enables the continuous tuning of the laser spectral bandwidth from $\sim30$ nm to $\sim300$ nm, delivering laser pulses with near Fourier Transform Limited (FTL) durations in the ranging from 23~fs to 3.5~fs. This remarkable feature makes it possible to understand the role of the pulse duration without having to introduce a frequency chirp in the laser pulses. In a way, this amounts to comparing different lasers systems with different bandwidths in a single experiment.

\begin{figure}[ht!]
\includegraphics[width=0.95\linewidth]{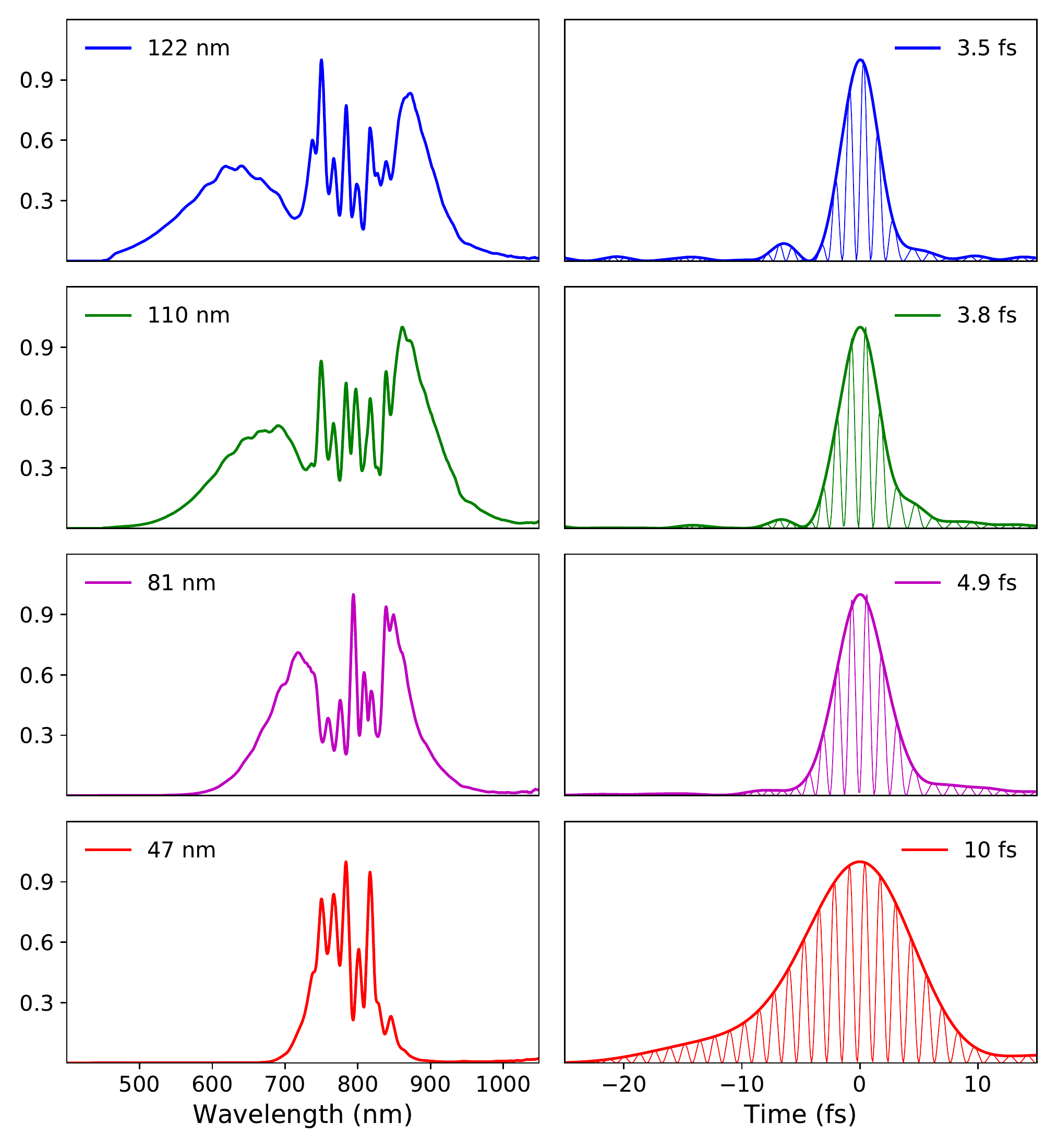}
\caption{Laser spectral (left) and temporal intensity (right) measured for different hollow core fiber pressures ${P_{HCF}=1100}$~mbar, $1000$~mbar, $700$~mbar, $250$~mbar (from top to bottom). Legends provide the RMS spectral width $\sigma$ (left), and FWHM duration $\tau_\text{fwhm}$ (right) for each  $P_{HCF}$. The thin curves represent $|E|^2$, and the thick curves the envelope. All curves are normalized.} \label{Fig1} 
\end{figure}

In a second study, we focus on stabilizing the LPA, which is of essential importance for actual applications of the electron beam. We show that controlling the injection using specially designed shocked gaz nozzles, and introducing a stabilization feedback system in the laser system greatly enhance the stability of the accelerator. Stable operation at 1\,kHz repetition rate over several continuous hours of operation is demonstrated \cite{rovige20}. The paper is organized as follows: section II describes the experimental and simulation methods used throughout this work. Section III shows the result of our extensive parametric study and proposes a discussion based on Particle-In-Cell simulations while section IV focuses on the stability study. Finally, section V summarizes and concludes.

\section{\label{sec:methods}Methods}

\paragraph{Laser system.} The experiments that follow were conducted using the Salle Noire kilohertz laser system at Laboratoire d'Optique Appliqu\'ee (LOA). The laser generates  ${\approx 10}$~mJ pulses with central wavelength ${\lambda_0=800}$~nm, and  full width at half maximum (FWHM) duration of  $23$~fs. The pulses are then post-compressed in a $2.5$~m-long hollow core fiber (HCF) filled with Helium gas \cite{boeh14,Ouille2020}. Varying the pressure of Helium in the fiber gives control over the spectral bandwidth and therefore over the pulse duration, which was tuned between $3.5$~fs and $23$~fs in the experiment, see \cref{Fig1}. The spectral width and the pulse temporal profile were measured in vacuum using the d-scan technique \cite{mira12}. The residual frequency chirp of the laser pulses was minimized by fine-tuning the dispersion using a pair of silica prism. In practice, this means that by tuning the pressure in the HCF fiber, we are able to vary the Fourier Transform Limited (FTL) pulse duration. Downstream, the laser was focused on target by, respectively for the the experiments of section \ref{sec:FTL}  and \ref{sec:OSS}, a f/2 and f/4 off-axis parabola to a ${3\times3}$~$\mu$m spot (FWHM) and to a ${6.2\times5.5}$~$\mu$m spot. In section, \ref{sec:FTL}, the energy on target is $\approx 2.5$~mJ, and the maximum laser amplitude in vacuum achieved for $3.5$~fs pulse was ${a_0\simeq1.5}$. Overall, the laser energy did not vary significantly ($<$2\%) when changing the pressure in the hollow core fiber. In section, \ref{sec:OSS}, the energy on target is $\approx 3.8$~mJ, the shortest pulse duration was $4.1$~fs and the maximum laser amplitude in vacuum was estimated to be ${a_0\simeq1}$.

\paragraph{Gas target.} For the LPA target in Sec.\,\ref{sec:FTL} we used a convergent-divergent nozzle with a $\SI{40}{\micro\meter}$ throat, $\SI{120}{\micro\meter}$ exit diameter which generated a continuous supersonic flow of molecular nitrogen ${N_2}$. The foot of the laser pulse is used to create the plasma by breaking the molecular and L-shell atomic bonds of $N_2$, thus releasing $10$ electrons per molecule, and providing a high density electron plasma. Additionally, When the laser field hits its peak,  ${a_0 \gtrsim 1}$, the laser pulse can ionize the K-shell and release two more electrons, thus triggering ionization injection \cite{pak10,mcgu10} (see also examples in the context of kHz LPA in \cite{guenot17, gustas18, faure19}). In Sec.\,\ref{sec:OSS} we used a $\SI{100}{\micro\meter}$-throat, $\SI{300}{\micro\meter}$-exit hole, one-sided shock (OSS) nozzle \cite{rovige20}, consisting of a supersonic nozzle at the end of which was added a $\SI{100}{\micro\meter}$ flat section on only one lateral side. This geometry creates an oblique shock \cite{Zucker2002} which generates a density bump followed by a downward transition at the beginning of a transverse path across the jet's flow. These jets were used for density gradient injection  \cite{bula98,bran08,gedd08,faur10,schm10}.

\begin{figure}[ht!]
 \centering
 \includegraphics[width=0.9\linewidth]{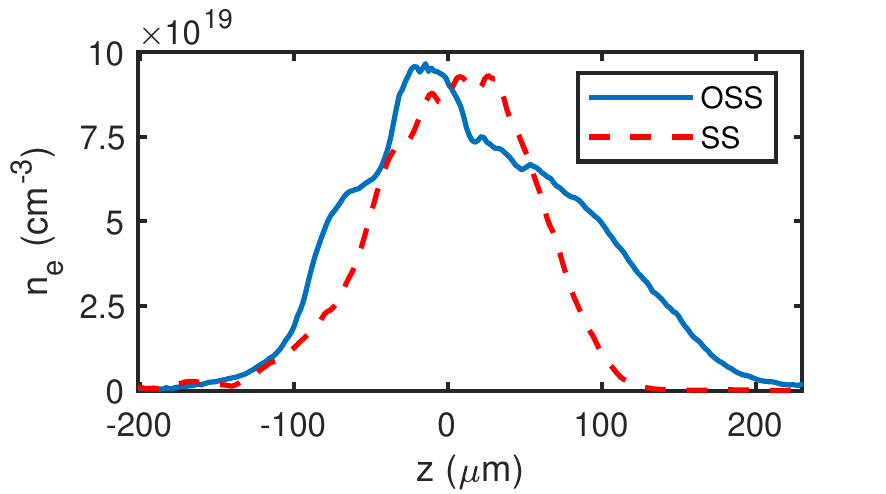}
 \caption{Measured plasma profile of the one-sided shock jet in nitrogen with a 22\,bar backing pressure (blue,solid) and plasma profile of the supersonic jet with a 25\,bar backing pressure and assuming L-shell ionization (red, dashed) at $\SI{150}{\micro\meter}$ from the nozzle.}\label{fig:OSS_density} 
\end{figure}

\paragraph{Diagnostics.} The gas and plasma profiles were characterized using a quadri-wave lateral shearing interferometer \cite{Primot1995,Primot2000} and the density maps were obtained via Abel inversion of the measured phase maps. For the supersonic jet used in Sec.\,\ref{sec:FTL}, we measured a quasi-gaussian density profile ${\propto \exp(-z^2/L_p^2)}$ with ${L_p=65}$~$\mu$m on the laser axis ($150$ $\mu$m from the nozzle). For backing pressures between 12~bar and 100~bar, the peak electron density varied from $4.2\times 10^{19}$~cm$^{-3}$ to $3.5\times 10^{20}$~cm$^{-3}$. The OSS jet used in Sec.\,\ref{sec:OSS} was operated with a backing pressure of 22 bar. The peak density was measured to be around $\SI{9.7e19}{\per\square\centi\meter}$ and the density after the shock was $\SI{7.3e19}{\per\square\centi\meter}$ , corresponding to a 25\% density drop with a transition width of $\SI{15}{\micro\meter}$. The measured plasma profile of both jets is displayed in Fig.\,\ref{fig:OSS_density}. 

The electron beam was characterized using a calibrated CsI(TI) phosphor screen, imaged onto a CCD camera. The laser beam was blocked by a thin aluminum foil in front of the detector, which also stopped electrons with energies lower than $\approx100$~keV. The energy of the electrons was measured with a removable spectrometer made of a $500$~$\mu$m pinhole and two permanent circular magnets. \par

\paragraph{Simulations.} Numerical simulations were performed using the particle-in-cell code FBPIC, which operates with quasi-cylindrical geometry \cite{lifs09}, and resolves electromagnetic equations using a pseudo-spectral algorithm \cite{Lehe:CPC2016, Andriyash:PoP2016}. The interaction domain was discretized in a cylindrical grid with cell-sizes $\Delta z \times \Delta r = 21\,\text{nm} \times 40\,\text{nm}$ in Sec.\,\ref{sec:FTL} and $\Delta z \times \Delta r = 27\,\text{nm} \times 70\,\text{nm}$ in Sec.\,\ref{sec:OSS}. Five azimuthal Fourier modes were used. The plasma was initially modeled as a neutral Nitrogen gas with 16 macro-particles per cell and with a density profile corresponding to measured experimental profile. The laser beam was initialized in the simulation using an antenna at the focal position, such that the spatial profile matches the one of the experiment. It was then back-propagated in vacuum to its initial position before focusing in the jet. Finally, the experimental measurements of the temporal laser field (see in \cref{Fig1}) were used as inputs for the simulations.

\section{\label{sec:FTL}Optimization and parametric study}

\begin{figure*}[ht!]
 \centering
 \includegraphics[width=0.95\linewidth]{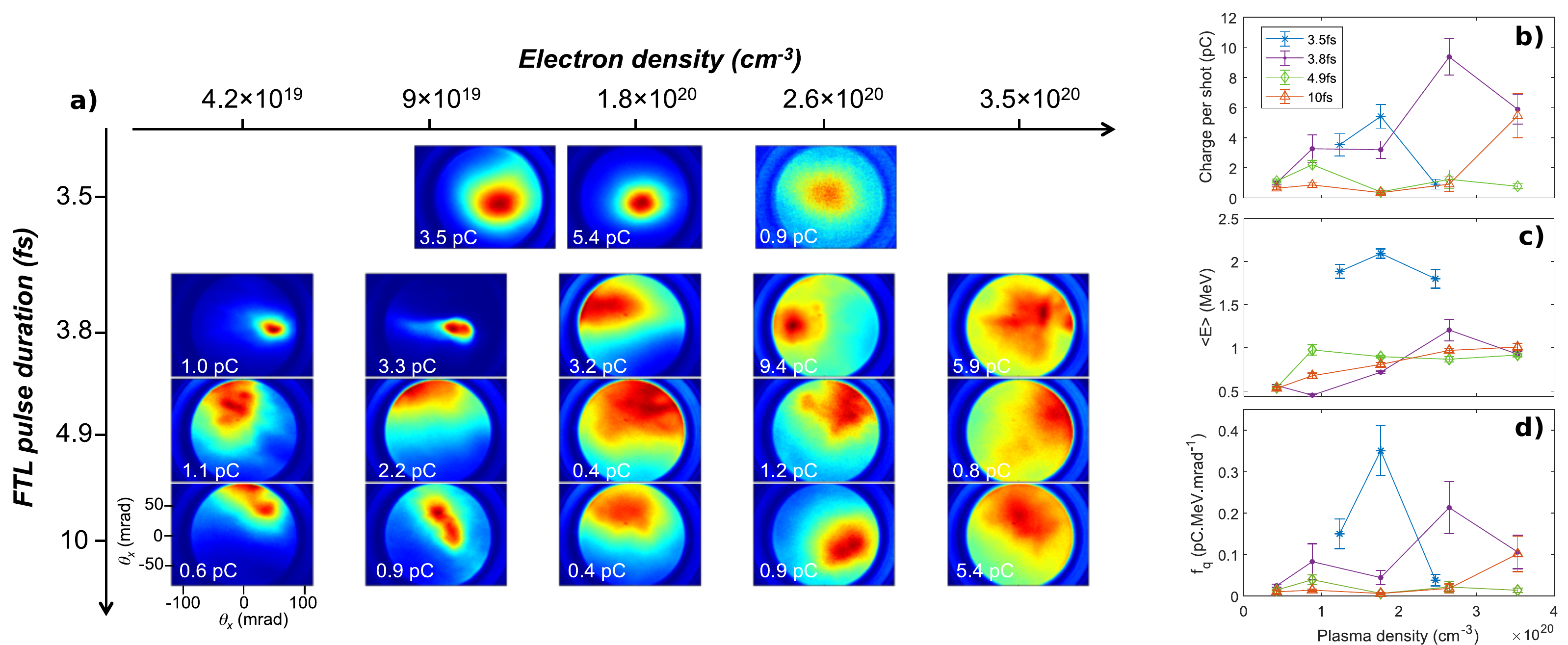}
 \caption{(a) Electron angular distributions for different plasma density and laser pulse FTL durations. Each image presents the average of 10 to 1000 shots, and the colorscale is then normalized to its maximum level. On the right, the following beam parameters are plotted as a function of plasma density, for different FTL pulse durations: charge per shot (b), mean energy of the accelerated electrons (c), and the quality factor as defined in the main text (d).The error bars represent the standard deviation from the mean value due to statistical fluctuations. } \label{ExperimentData} 
\end{figure*}

\paragraph{Results.} We now present a parametric study where we measured the electron signal for different plasma densities and laser pulse durations. In the rest of the paper, we refer to the FWHM pulse duration as the Fourier transform limited (FTL) pulse duration to clearly indicate that the pulse duration is changed by tuning the spectral bandwidth and not by introducing a chirp. For each point $(\tau_\mathrm{fwhm}, n_e)$, a d-scan measurement was performed in order to ensure that the laser pulse is optimally compressed. The raw data for the electron beams are presented in \cref{ExperimentData}a, where each image was obtained by averaging over 10 to 1000 laser shots, depending on the signal level. Electron beams were never observed for the case of 23~fs FTL duration so that this case is not represented. In all other cases, an electron beam was obtained while its charge and divergence could vary greatly depending on the exprimental parameters. 

A global analysis of electron beam data is presented in \cref{ExperimentData}b-d, where we plot the mean value, the charge (\cref{ExperimentData}b) and the average electron energy (\cref{ExperimentData}c) as a function of laser pulse FTL duration and electron plasma density. Each point is averaged over 20 acquisitions, and thus represents an average over 200 shots minimum (high signal case) and 20000 shots maximum (for the low signal cases). The vertical error bars represent the standard deviation from the mean value estimated from the 20 acquisitions. To quantify the LPA performance, we define  the quality factor, $f_q = {\langle E\rangle\, Q}/{\sqrt{\sigma_x\sigma_y}}$,  shown in \cref{ExperimentData}d. In this formula, $Q$ is the total charge per shot, $\langle E\rangle$ the average energy of accelerated electrons, and $\sigma_x$, $\sigma_y$ are the RMS angular divergences of the beam along $x$ and $y$ directions. Clearly, this quality factor favors electron beams with high energy, high charge and narrow divergence. The first striking result is that the highest quality factor is obtained for the shortest pulses, 3.5~fs, at a specific resonant density.

\begin{figure}[ht!]
 \centering
 \includegraphics[width=\linewidth]{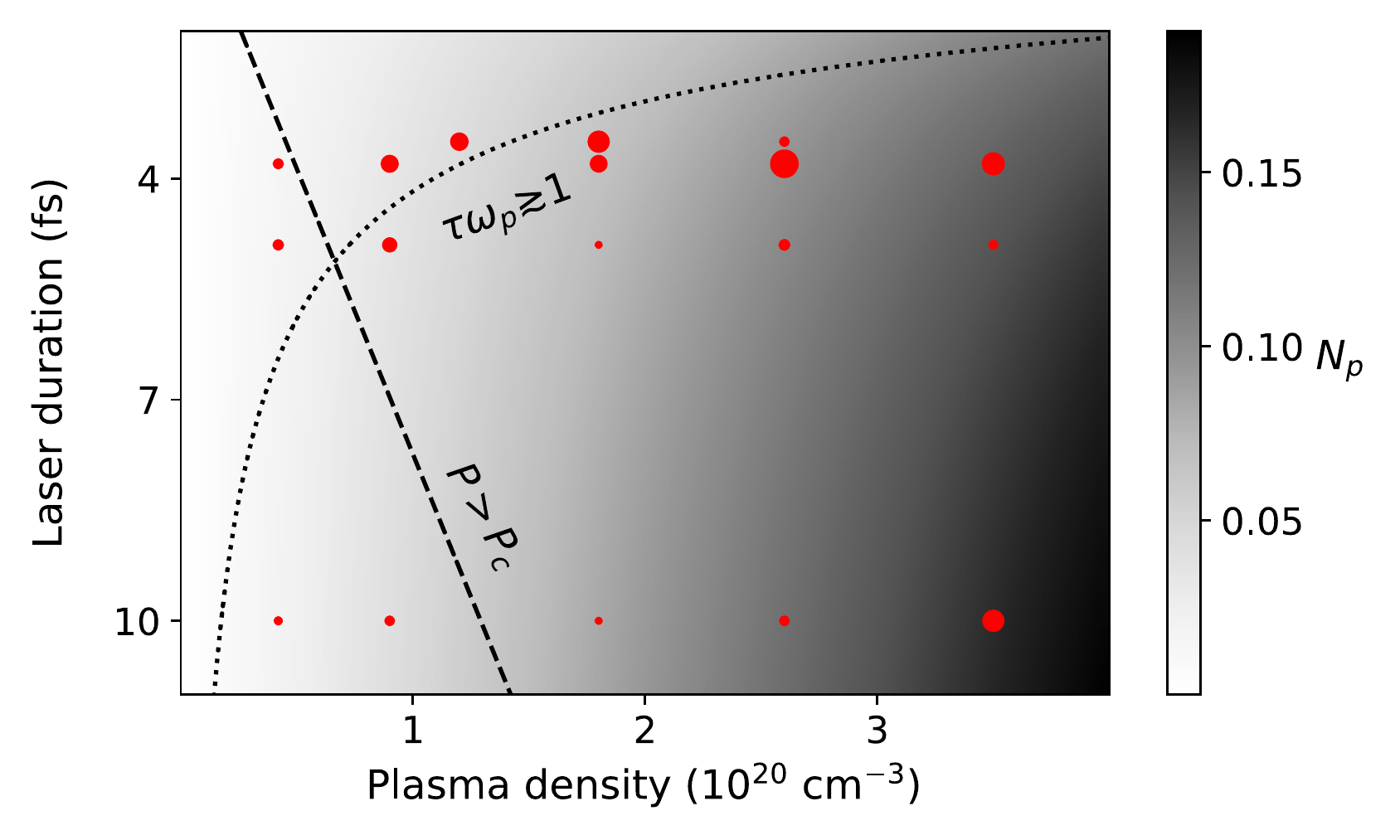}
 \caption{
 Parameter space of the kHz LPA. Measurements are represented by red circles with sizes proportional to the beam charge.  Dashed and dotted lines represent conditions in \cref{eq:matching}, and the gray colormap corresponds to the normalized plasma density $N_p = n_p / \gamma_\perp n_c$.
 }\label{fig:discuss} 
\end{figure}

\paragraph{General considerations.} Let us now consider correlations between experimental parameters and LPA performance. In \cref{fig:discuss}, we plot the same parameter space as in \cref{ExperimentData}a, and show the measurements using red circles whose size represent the total beam charge. On the same graph, the conditions for relativistic self-focusing, and longitudinal matching of the laser pulse with the plasma are also represented. We recall these conditions:
\begin{equation}\label{eq:matching}
P_0\gtrsim 17 \, n_c/n_p\;\text{[GW] (a), and } c\tau\simeq c/\omega_p \text{ (b)}\,,
\end{equation}
where $n_c=1.1\, \lambda_0^{-2}\, 10^{21}$~cm$^{-3}$ is the plasma critical density for laser wavelength $\lambda_0$ (here in micrometers), and ${\omega_p = c\sqrt{4\pi r_e n_p}}$ is the plasma frequency with the plasma density $n_p$, and the classical electron radius $r_e$. Note that the pulse duration $\tau$ in Eq.~(\ref{eq:matching}b) is defined as the RMS of the intensity profile and corresponds to $\tau=\tau_\text{fwhm}/2.355$. For these estimates, it was assumed that  the laser energy is 2~mJ and the focal spot size $R_\text{fwhm}=3$~$\mu$m. 

In \cref{fig:discuss}, we plot Eq.~(\ref{eq:matching}a) and Eq.~(\ref{eq:matching}b) with the dashed and dotted  curves respectively. Firstly, one may note that below the relativistic self-focusing threshold, the accelerated charge is low, indicating that self-focusing is required to reach the high intensity required for stable self-injection. Most of the high-charge cases correspond to above-critical laser powers and pulse duration close to the plasma wave periods -- cases of $3.5$~fs and $3.8$~fs laser pulses. Figure \ref{ExperimentData}a shows that for the $3.8$~fs pulse, as the plasma density grows, the injected charge increases, and so does the beam angular divergence. The charge increase can be explained by stronger self-focusing and also by the laser group velocity slowdown in the plasma, $v_g/c \simeq 1 - n_p/(2 \gamma_p n_c)$, where ${\gamma_\perp\simeq\sqrt{1+a_0^2/2}}$ is the relativistic factor of laser-driven electron fluid. This plasma wave slows down for the higher $n_p$, which facilitates co-phasing of electrons with the wake, as required for injection. Beam divergence is determined by the spread of electron transverse momenta acquired during the injection. Naturally, this initial divergence scales with the plasma focusing force growing as $F_\perp\propto n_p$.

\begin{figure*}[t!]
 \centering
  \includegraphics[width=0.99\linewidth]{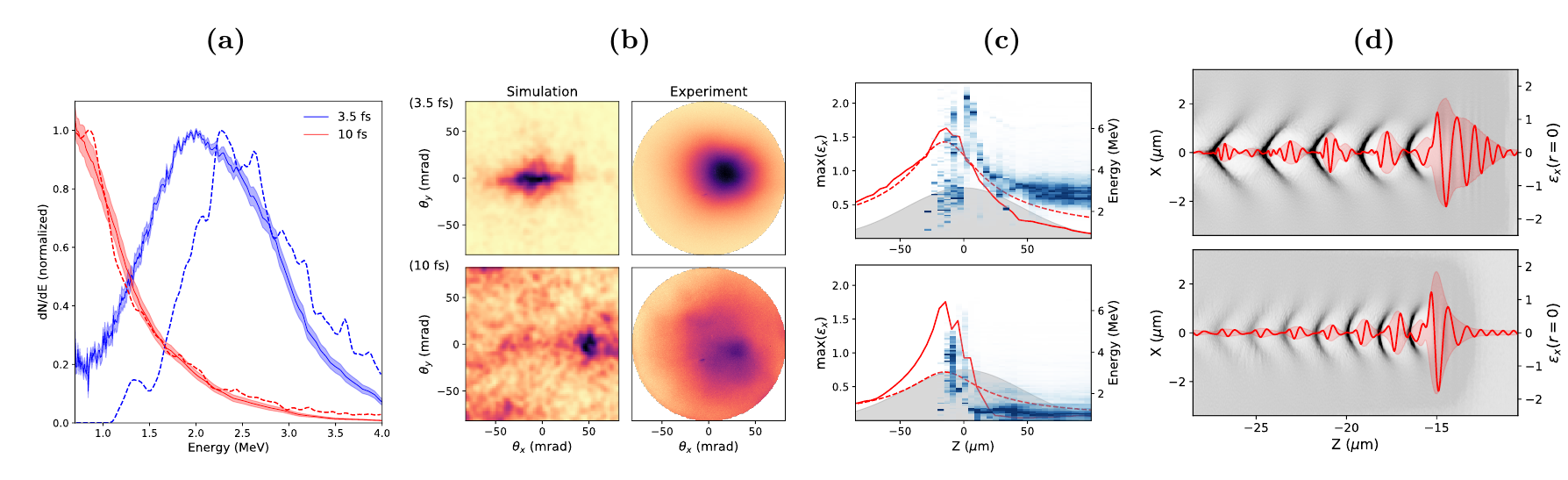}
 \caption{Measured and modeled electron beams with $3.5$~fs laser pulse and $n_p=1.8 \times 10^{20}$~cm$^{-3}$ (blue in (a), top in (b,c,d)), and with $10$~fs pulse and $n_p=3.5 \times 10^{20}$~cm$^{-3}$ (red in (a), bottom in (b,c,d)). (a) Measured electron spectra (solid curves shaded with standard deviation) and modeled (dashed curves). (b) Measured (right) and simulated (left) angular electron distribution. (c) Simulated evolution of the peak laser field $\varepsilon_x = eE_x/m_e c\omega_0$ (red curve) and the electron energy spectrum normalized at each time-step (blue) (blues colors); the dashed red curve is the in-vacuum laser propagation, and the gray area represents the plasma density profile in arbitrary units. (d) On-axis laser field (red curve), its temporal envelope (red filled area), and electron density normalized to $15n_p$ (gray scale) at the laser focal position, just before injection.
 }\label{fig:PIC} 
\end{figure*}

The longer laser pulses (10~fs case) clearly do not fulfill the longitudinal resonance condition Eq.~(\ref{eq:matching}b), and its $a_0\approx 1$ is too low to trigger self-injection. These longer pulses need to evolve in the plasma, via the self-modulation and relativistic self-focusing instabilities to efficiently excite a plasma wave. For the considered parameters, these processes are strongly nonlinear and can be quantified by the normalized plasma density, $N_p = n_p / \gamma_\perp n_c$, an equivalent to the similarity parameter in Ref.\,\cite{Gordienko:PoP2005}. For higher $N_p$, one may expect the plasma nonlinearities to dominate the physics. In \cref{fig:discuss}, $N_p$ is plotted in gray, and we see that among the unmatched cases  ($4.9$~fs and $10$~fs), efficient electron acceleration occurs preferentially for the case when the laser is sufficiently long, $\tau\omega_p\gg1$, and when plasma parameter reaches $N_p\gtrsim1/6$. The first condition is favorable for the self-modulation instability while high $N_p$ favors nonlinear plasma effects.

\paragraph{Simulations.} In order to understand the transition between these regimes, we performed PIC simulations for two extreme cases -- the shortest (${3.5}$~fs) and the longest (${10}$~fs) laser pulses with their corresponding plasma densities ${n_e =1.8 \times 10^{20}}$~cm$^{-3}$ and ${3.5 \times 10^{20}}$~cm$^{-3}$, respectively. These two cases correspond to (i) the resonant case, giving peaked electron energy and narrow divergence beams, and  (ii) the self-modulated case, giving high charge and high divergence beams. In the simulations, the laser energy was adapted to match the experimental results, giving $2$~mJ for the $3.5$~fs case, and 1.5~mJ for the $10$~fs case.

In \cref{fig:PIC}(a,b) we compare the spectral and angular electron distributions obtained in experiment to the simulation results. For the chosen parameters, the simulated beam features are in a good agreement with the experiment. We see that the shortest pulse provides a collimated electron beam with a peak in spectrum around $2-2.5$~MeV, while in the $10$~fs case, the beam is rather divergent and has a thermal spectrum with an equivalent ``temperature'' ${T_{e}\approx 700}$~keV. In the experiment, the total charge was measured within a 75 mrad aperture, and was found to be 5.4~pC in both cases. In simulations, considering the same aperture, the beam charges are 4.6~pC (22~\% from K-shell ionization) and 9~pC (16~\% from K-shell) in the $3.5$~fs and $10$~fs pulse cases respectively. However, in the latter case, the modeled beam divergence was $\sim 200$~mrad (fwhm) and the total charge reached $48$~pC (mainly from the self-injection) in that larger aperture.

The dynamics of the electron spectra and laser peak field are depicted in \cref{fig:PIC}(c). In the $3.5$~fs case, the laser experiences moderate self-focusing and produces electron injection and acceleration, which brings electrons to high energies $\lesssim 8$~MeV. Acceleration is followed by the \textit{electron dephasing}, i.e. electrons overrun the wake. When particles reach the front half of the bubble, the plasma field tends to decelerate them down to the energy level given by the wake phase velocity, i.e. the laser group velocity: ${\epsilon_w=m_ec^2(\gamma_g-1) \simeq m_ec^2 (N_p^{-1/2}-1)}$. For the upper case of \cref{fig:PIC}(c), this energy is ${\epsilon_w\simeq 1.5}$~MeV, but laser diffraction stops the deceleration when $a_0\lesssim 0.5$, and leaves the electron spectrum with a peak around $2.5$~MeV. In the second case (\cref{fig:PIC}(c) bottom), the laser gets strongly focused and modulated, $a_0\gtrsim 1.7$, which triggers electron injection. Although self-focusing greatly enhances the laser field, the interaction length and the acceleration length are both strongly reduced because of strong pump depletion and strong defocusing. In this case, electron dephasing is also very fast (distance of few $\mu$m), so that electrons are continuously accelerated-decelerated, which produces a thermal electron spectrum around, ${\epsilon_w=700}$~keV.

Laser self-modulation is essential to achieve laser-plasma acceleration for the longer pulses, $\omega_p\tau \gg 1$, which in our study occurs with a $10$~fs pulse duration. In \cref{fig:PIC}(d), we see that in both short and long cases, the pulse profile at its focal position acquires a peak located in the first bucket of the plasma wave. This ``driver'' peak is shaped by the plasma resonance, and its duration, measured with a Gaussian fit, matches exactly the condition Eq.~(\ref{eq:matching}b), giving $\tau_\text{fwhm}=3.1$~fs and $2.1$~fs for ${n_e =1.8\cdot 10^{20}}$~cm$^{-3}$ and ${3.5\cdot10^{20}}$~cm$^{-3}$ respectively. Surprisingly, the physics of these two cases is thus relatively similar. The main difference is that in the high density case, the nonlinearities are more violent and the pump depletion length is shorter, which greatly limits the acceleration length.

The conclusion of these simulations is that there is still room for improvement: higher electron energy, up to 8-10 MeV, might be achieved if the plasma length could be made shorter so that electrons would not dephase. This could be done with even shorter jets for example.

\section{\label{sec:OSS}Stable continuous operation with a one-sided shock jet}

\begin{figure*}[t!]
 \centering
  \includegraphics[width=0.9\linewidth]{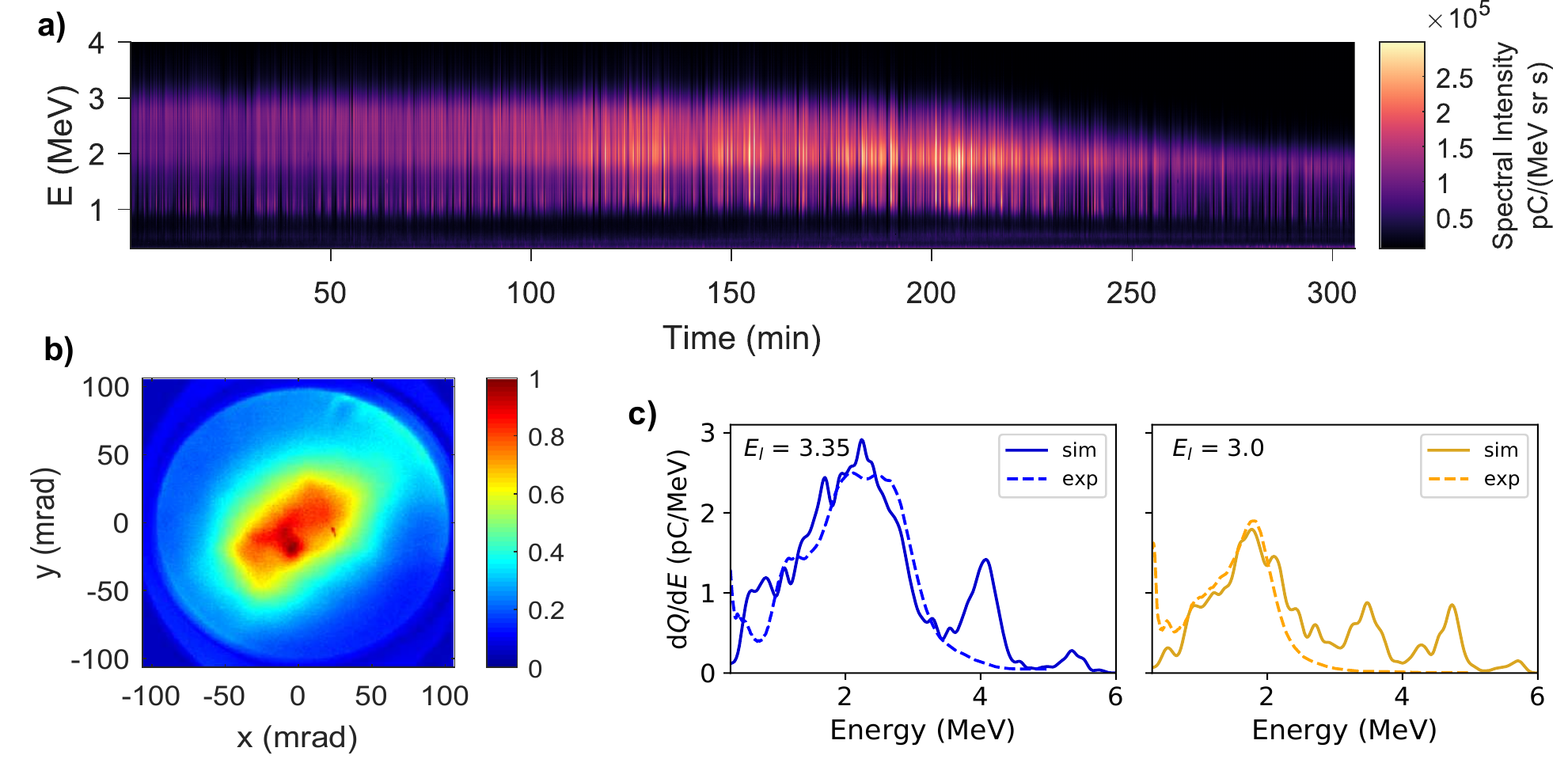}
 \caption{a) Electron spectra monitored during the 5h run. Each spectrum is  averaged on 100 shots. b) Typical beam measured before the run. The mean charge of 2.6\,pC/shot is an average on 200 shots . c) Simulated (solid line) and experimental (dashed line) electron spectra at the beginning of the experiment (blue, simulation with an adjusted 3.35\,mJ energy) and at the end of the experiment (yellow, simulation with 3.0\,mJ ) }\label{fig:elecstab} 
\end{figure*}

\paragraph{Results.}Even though the configuration depicted in the previous section using a  $\mathrm{120\,\upmu m}$ supersonic jet, yielded good quality beams in the short pulse case, the mid-to-long term stability, i.e, keeping the beam consistent for more than a few minutes, has proven to be a challenge. Moreover, the day-to-day reproducibility in this configuration was not satisfactory, as the accelerator performances could considerably vary from one day to another, while known laser and plasma parameters would remain similar. In order to solve these issues, a new $\mathrm{300\,\upmu m}$ one-sided shocked supersonic jet providing a density downward transition at the beginning of the plasma was used (see \cref{fig:OSS_density}). The downward density transition is known to facilitate trapping and to enable controlled injection in the density ramp\cite{bula98,bran08,gedd08,faur10,schm10}. The target characteristics and experimental configuration are thoroughly described in \cite{rovige20}. We draw attention to the fact that the gas target dimensions are larger than in the previous experiment with a supersonic nozzle, mostly due to manufacturing constraints. Therefore the focal length was also increased from $f=50$~mm to $f=100$~mm in order to increase the Rayleigh length for a better match with the plasma profile. \par
A second important improvement was the introduction of feed-back loops so as to stabilize the performance of the laser system. In particular, we found that it was critical to stabilize the laser beam pointing at the level of the last focusing optics. This permitted to maintain the position and the spatial quality of the focal spot throughout the experimental run. These improvements allowed us to perform a 5\,h-long continuous and stable hands-off operation of our accelerator with an accelerated charge of $ 2.6\pm 0.6$\,pC/shot measured beforehand, with beam divergence of $80 \times 75\pm 8\times 9$\,mrad. The electron spectrum was monitored constantly during the run, and showed a remarkable stability during the first 150\,min with a peaked spectrum at 2.5\,MeV. The results are displayed in Fig.\,\ref{fig:elecstab}. It appears that after 150\,min the high-energy part of the spectrum started to erode, and at the end of the run, the peak of the spectrum was displaced from 2.5\,MeV to 1.9\,MeV. This decrease in energy of the electrons can be associated to a continuous decrease in the monitored laser energy of $11\%$ during the experiment. (the cause of which was later attributed to a series of damages on the laser chain). Still, the accelerator consistently produced electron beams over 5 hours, with peaked spectra and energies in the range 1-3\,MeV. This represents a dramatic improvement compared to previous experiments with supersonic nozzles in which the longevity of such beams in hands-off operation would be of a few minutes. In addition, even though it led to a loss of performance, the decrease in laser energy highlights the robustness of the density downramp injection method, since this advance in long-term stability was achieved despite the significant variation of an important parameter.\par

\begin{figure*}[t!]
 \centering
  \includegraphics[width=0.99\linewidth]{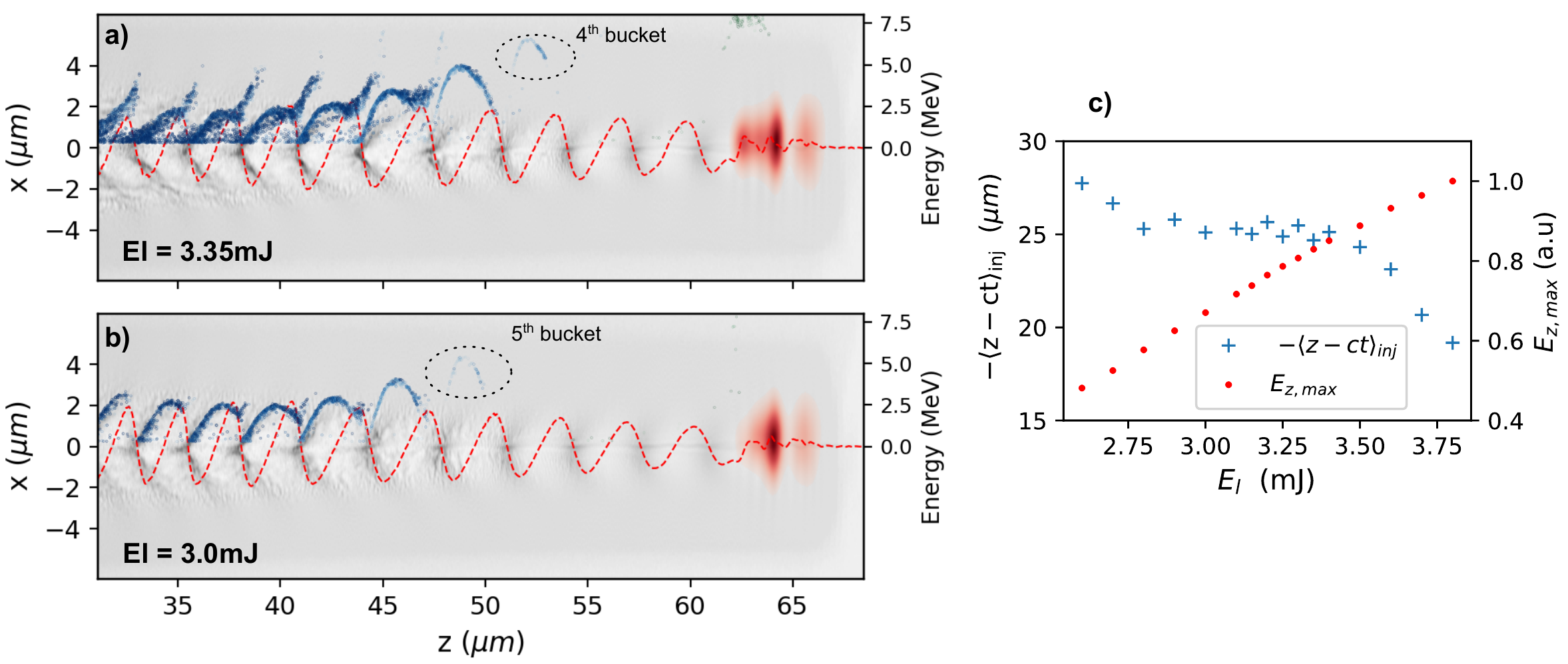}
 \caption{Electron density (gray scale) in $z{-}x$ plane, and in phase space $z{−}\epsilon_z$, where $\epsilon_z$ is the energy neglecting the contribution of momentum along x and y, for L-shell and K-shell electrons (blue and green respectively), longitudinal electric field ($E_z$) (red dashed curve), and laser envelope (red shaded area) for the a) : $E_l=$3.35\,mJ case and  b) $E_l=$3.0\,mJ case.  c) Mean position of injection behind the laser pulse and maximum longitudinal electric field (normalized to the value for $\mathrm{E_l=3.8\,mJ}$) according to the laser driver energy}\label{fig:simushock} 
\end{figure*}

\paragraph{Simulations.} Particle-in-cell simulations with experimental parameters as input were performed in order to reproduce these results, and gain more insight into the injection and acceleration processes, with a specific focus on the dependence on the laser driver energy. To do so, 16 simulations were performed, varying $E_l$ in the range 2.6-3.8\,mJ. Figure \ref{fig:elecstab}c represents the experimental and simulated spectra for laser energy corresponding to the beginning and the end of the experimental run. The energy in the corresponding simulations had to be slightly reduced to 3.35\,mJ and 3.0\,mJ (maintaining the 10\% drop) to match the experimental results. The simulations reproduce quite well the electron energy loss concomitant to a  $10\%$ decrease of the laser energy, even though some higher energy electrons (small peaks at $E>3$\,MeV) were not detected in the experiment. The charges in the simulations are 6.2\,pC for the 3.35\,mJ case and 3.3\,pC with the 3.0\,mJ pulse. These simulations indicate that injection does occurs in the density transition region, supported by self-focusing which is triggered in the density bump in the shock region, see Ref.\cite{rovige20}. Electrons from ionization injection represent only $5\%$ of the total charge in the 3.35\,mJ case, and lead to a small peak around 7.5\,MeV in the spectrum. This peak was not detected during the experiment and is not displayed in Fig.\,\ref{fig:elecstab}c for the sake of clarity. \par

In fig.\,\ref{fig:simushock}a-b, we represent, for two laser energies, the plasma density at the end of the downramp, the injected electrons in phase space and the on-axis $E_z$ field. In both cases, L-shell electrons are trapped in several buckets, starting quite far behind the laser pulse, respectively from the $\mathrm{4^{th}}$ and the $\mathrm{5^{th}}$. This behavior is found in all the simulated cases: it is found that the first bucket in which electrons are injected moves backward. It goes from the  $\mathrm{3^{rd}}$ bucket at $E_l=3.8$\,mJ to the $\mathrm{7^{th}}$ bucket at $E_l=2.6$\,mJ. This can be explained by considering the evolution of the wake phase velocity in a downward density transition \cite{bran08} :

 \begin{equation}\label{eqn:vp}
v_{p} = c\frac{1}{1+(z-ct)\frac{1}{k_p}\frac{dk_p}{dz}}\,,
\end{equation}
With $k_p(z)=\sqrt{4\pi r_e n_e(z)}$ the local plasma wave number where $r_e$ is the classical electron radius. According to Eq.\,\ref{eqn:vp} the wake phase velocity decreases with the laser co-moving coordinate $\mathrm{(z-ct)}$, The immediate consequence of this is that $v_{p}$ is smaller in buckets further behind the laser therefore easing trapping in these buckets. When lowering the laser energy, the amplitude of the wake decreases (see Fig\,\ref{fig:simushock}c) and the first bucket where the wavebraking condition is satisfied moves further behind, leading to the behaviour of moving back of the injection observed in Fig\,\ref{fig:simushock}a,b. This is supported by the evolution of the mean position of injection in the co-moving frame, according to the laser initial energy plotted in Fig\,\ref{fig:simushock}c, where the mean position of injection $\langle z-ct\rangle_{inj}$ goes from $\SI{-19}{\micro\meter}$ at 3.8\,mJ to $\SI{-28}{\micro\meter}$ at 2.6\,mJ\par

It is clear from Fig.\,\ref{fig:simushock}a,b and the typical parabolic shape of the electrons in phase space that electrons energy is limited by dephasing. As the phase velocity $v_p$ decreases with the laser co-moving coordinate in the density transition region, so does the associated Lorentz factor $\gamma_p=(1-v_p^2/c^2)^{-1/2}$, and so does the dephasing length $L_{deph}\propto \lambda_p\gamma_p^2$. This leads to smaller energies further behind the laser. This can be clearly observed in both panels a) and b) of Fig.\,\ref{fig:simushock} where the maximum energy of each parabola in phase space is gradually smaller than the one before. The decrease in electron energy with the laser energy can therefore be explained by the backward shift of the injection position observed in Fig.\,\ref{fig:simushock}c, leading to a globally smaller dephasing length.\par

\begin{figure}[ht!]
 \centering
  \includegraphics[width=0.99\linewidth]{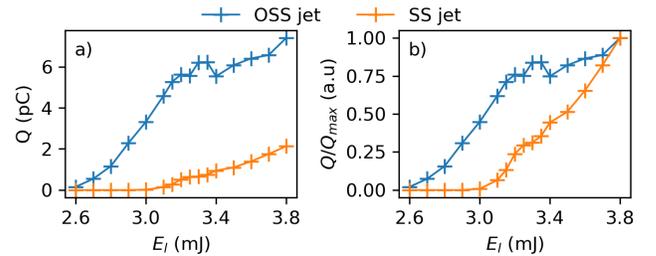}
 \caption{a) Absolute accelerated charge in simulations and b) The same curves normalized to the charge obtained at $E_l=3.8$\,mJ to show the superior stability of the OSS jet against laser energy fluctuations.. }\label{fig:chargeOSS} 
\end{figure}

In Fig.\,\ref{fig:chargeOSS} we compare the accelerated charge according to the input laser energy in the simulations, between both shocked and supersonic gas profiles  with otherwise the exact same set of parameters (the laser is focused at the maximum of the gaussian plasma profile in the supersonic case). This shows that for the considered parameters, the shocked jet yields a significantly higher charge for all the energies (6.2\,pC vs 0.8\,pC at $E_l=3.35\,mJ$) due to the eased injection in the density transition region. Moreover, the normalized curves highlight quite well the enhanced stability obtained with the OSS jet, as in this case, the decrease in charge with the laser energy is, in a first instance, much more gentle than with the supersonic jet.\par
From this analysis, we can determine that the use of shocked nozzles for high-repetition rate LWFA presents several advantages compared to supersonic jets, notably in term of stability and accelerated charge. However, the yielded beams significantly more divergent, and the temporal profile of electron bunches stemming from multi-buckets injection in the gradient is necessarily elongated. In order to improve this, a steeper gradient combined with a higher laser intensity would allow trapping in the first bucket. The simulations demonstrate that the electron energy is largely limited by the short dephasing length that derives from the high plasma density necessary to achieve laser self-focusing. A higher pulse energy would allow us to work with lower plasma density, but this represents a challenging technological advance. A smaller shocked gas jet, of dimensions comparable to the supersonic gas jet used in Sec.\,\ref{sec:FTL}, more adapted to our typical acceleration lengths, would probably yield better results, and while still technologically demanding, such small asymmetric jets can be manufactured and will be used in future experiments.       
 
% \begin{equation}\label{eqn:ldeph}
%L_{phi} = \frac{\uplambda}{2}\left( 1+\frac{1}{(z-ct)\frac{1}{k_p}\frac{dk_p}{dz}}%\right) \,,
%\end{equation}

\section{\label{sec:conclusion}Conclusion}

We have studied laser-plasma acceleration with a few-cycle few-mJ laser in a wide range of laser pulse durations and plasma densities. We have identified two regimes for generating multi-pC beams. In the first regime, the resonance condition is fulfilled as the laser pulse duration matches the plasma period (3.5~fs laser pulses and $n_p \sim 1.8 \times 10^{20}$~cm$^{-3}$). This leads to collimated electron beams with a 2-3 MeV peaked energy spectrum. In the second case, a longer duration pulse, 10~fs, is strongly self-modulated and self-compressed in a higher density plasma ${n_p \sim 3 \times 10^{20}}$~cm$^{-3}$. This leads to electron beams with a higher charge and divergence, but with a thermal energy distribution extending to a few MeV energies. Although the resonant regime leads to higher quality electron beams, the high divergence and high charge regime is relevant for irradiation experiments where high radiation doses are required. Recently, we have operated our kHz laser-plasma accelerator in this regime in order to perform a radiobiology experiment in which high doses were deposited in cancerous cells\cite{cavallone20}.

Moreover, using a new one sided shock nozzle, we have achieved stable, continuous and hands-off operation of our LPA for more than 5 hours, accumulating more than $18\times10^{6}$\, consecutive shots, with pC charge and few-MeV electrons. This represents a significant advance toward the use of LPA as a source for applications. 

Numerical simulations, performed in a variety of cases, highlight the fact that our accelerator is limited by dephasing.  Therefore, decreasing the interaction length in order to avoid electron dephasing will be a requirement for increasing the energy to 10~MeV and beyond. Additionally, using lighter gases is also expected to increase the electron energy as was recently demonstrated \cite{salehi20}, which may also open new opportunities. 

\section*{\label{sec:acknowledgments}Acknowledgments}
We acknowledge  Laserlab-Europe, JSPS KAKENHI Grant Number 19H00668, H2020 EC-GA 654148 and the Lithuanian Research Council under grant agreement No. S-MIP-17-79.

\section*{\label{sec:Data availibility}Data availability}
The data that support the findings of this study are available from the corresponding author upon reasonable request.

%\bibliography{jerome_1220,suppl-bib}

%\bibliographystyle{aipnum4-1}
%merlin.mbs aipnum4-1.bst 2010-07-25 4.21a (PWD, AO, DPC) hacked
%Control: key (0)
%Control: author (8) initials jnrlst
%Control: editor formatted (1) identically to author
%Control: production of article title (-1) disabled
%Control: page (0) single
%Control: year (1) truncated
%Control: production of eprint (0) enabled
%

\end{document}